\documentclass[a4paper,11pt]{article}
\usepackage{pos}

\title{
New $Z_c$ spectroscopy or Landau singularities?\\ Searches in Heavy Ion Collisions
could help decide.
}

\author*[a]{Felipe J. Llanes-Estrada}
\author[b]{Luciano M. Abreu}

\affiliation[a]{Univ. Complutense de Madrid, Dept. F\'{\i}sica Te\'orica, Plaza de las Ciencias 1, 28040 Madrid, Spain}

\affiliation[b]{Instituto de F\'isica, Universidade Federal da Bahia, Salvador, Bahia, 40170-115, Brazil}

\emailAdd{fllanes@fis.ucm.es}
\emailAdd{luciano.abreu@ufba.br}

\abstract{
A constellation of quarkonium-mass peaks has been reported in the last decade, opening what could be an entire new spectroscopy of nuclear-physics like complexity. Salient  
among these structures are the $Z_c$, much analyzed at BESIII as nicely summarised by J.~Zhao at this conference. Their stature as new hadron states has often been challenged by proposals that the peaks arise from triangle singularities instead, 
because they appear in three-body reactions and often satisfy the special kinematic conditions of the Coleman-Norton theorem. \\
To experimentally decide, we should like to resort to different reactions erasing the kinematic coincidence, and see whether each peak survives. We observe that the most universal ``eraser'' is the medium created in Heavy Ion Collisions, because its high temperature often affects the masses and widths of the particles participating in the reaction: if additionally the triangle reaction is completed during the lifetime of the fireball, our computations in thermal field theory show that the triangle singularities do not survive the hot hadron phase (much less the hotter quark-gluon plasma). On the contrary, ordinary quarkonia are affected but often survive the fireball and molecular-like states (such as the antinuclei $^2\overline{H}$, $^3\overline{H}$ or $^4\overline{He}$) are also routinely detected. \\
We have provided several examples, and here investigate  the exotic $Z_{c}(4020)^\pm\to \pi^{\pm} h_c $ peak (produced against a $\pi^\mp$ in $e^+e^-$ collisions).
If, as has been proposed during the debates, it is produced or enhanced by a $D_{1}  D^{*} D^*$ triangle loop, the estimate here reported suggests its disappearance
near the critical temperature of the confinement/deconfinement phase transition, so any future detection in  heavy-ion collision data would suggest it to be a real hadron, with its absence leaning towards the singularity interpretation.
}

\FullConference{%
  *** Particles and Nuclei International Conference (PANIC2021), ***\\
  *** 5-10 September 2021 ***\\
  *** Online conference, organized by  LIP, Laboratory for Instrumentation and Experimental Particle Physics, and FCUL, Faculty of Sciences of the University of Lisbon (Portugal). *** 
}

\begin{document}

\maketitle

We have recently suggested~\cite{Abreu:2020jsl} that triangle singularities, engendered by kinematic effects complying with the Coleman-Norton theorem, might be strongly affected by the medium in a heavy-ion collision environment. The effect takes place only under the following conditions: the  triangle reaction must be completed in the lifetime of the fireball; and the mass or width of the particles in the triangle diagram must be sufficiently altered from their vacuum values. This may be of much importance for the new hadron spectroscopy, for example the $Z_c$ states in table~\ref{tab:zc}, since the medium can act as a filtering mechanism distinguishing kinematic accidents from real hadrons. 

\begin{table}[h]
    \centering
   \caption{Exotic $Z_c$-family peaks reported in $e^-e^+$ collisions at BES-III as summarized by J.~Zhao at the conference. All of them appear in two-body subsystems (underbraced, with conjugate modes not written down) of three-body final states, so that they might be influenced by hadron triangles~\cite{Bugg:2011jr}. One could add various states such as $Z_c(4200)$ or $Z_c(4430)$ seen by Belle, but this sample suffices to exemplify.}
    \label{tab:zc}
    \begin{tabular}{|c|cc|cc|c|} \hline \hline
Peak & $Z_c(3900)^{\pm/0}$ & $Z_c(3885)^{\pm/0}$ & $Z_c(4020)^{\pm/0}$ & $Z_c(4025)^{\pm/0}$ & $Z_{cs}(3985)^-$  \\ Final state & $\pi \underbrace{\pi J/\psi}$ & $\pi \underbrace{D\bar{D}^*}$ & 
    $\pi \underbrace{\pi h_c}$ & $\pi \underbrace{D^*\bar{D}^*}$ & $K^+ \underbrace{(D_s\bar{D}^*)^-}$  \\  
Triangle  & $D^*D^*D^0$ \cite{Wang:2013cya}  & same? & $D_{1/2}D^* D^*$ \cite{Liu:2014spa}  & same? & $D_{s2} D_s^* D^0$ \cite{Ikeno:2020mra} \\ \hline \hline
    \end{tabular}
\end{table}

Aside from the examples analyzed earlier~\cite{Abreu:2020jsl}, we have recently examined whether heavy ion collisions can make a statement about the nature of two of those new exotic states:  the $Z_c(3900) $ state seen in $Y(4260)\to J/\psi\pi\pi$ decays~\cite{Abreu:2021xpz}, and the $Z_{cs}(3985)$ structure newly observed by the BESIII Collaboration~\cite{Llanes-Estrada:2021jud}.
We here present another emblematic illustration in this series of studies on the interpretation of the exotic hadronic states, turning to the charged $Z_{c}(4020)$ state, the third entry in table~\ref{tab:zc}.

This state has also been observed by the BESIII Collaboration, in the $\pi^{\pm} h_c $ mass spectrum of the reactions $e^+e^- \rightarrow \pi^+ \pi^- h_c $, at center-of-mass energies from 3.90 GeV to 4.42 GeV~\cite{BESIII:2013ouc}. Its charged nature and intriguing properties make it a candidate exotic hadron with a minimum tetraquark $c \bar{c} u \bar{d}/d \bar{u}$ content. A lot of work has been devoted to clarify its underlying structure as a meson molecule, a compact tetraquark or due to triangle production effects, but a comprehensive understanding is still missing (we refer the reader to Ref.~\cite{Sakai:2021qrg} and references therein for a more detailed discussion). 

The interpretation associating it to some type of Landau singularity can be tested against a hadron interpretation, following our aforementioned studies,  in  heavy-ion collisions. There, these kinematic effects are erased by the medium at temperatures close to the critical temperature at which the confinement/deconfinement phase transition occurs as we here show. Therefore, any possible peak that would be reported in future heavy-ion collision data that might be interpreted as the $Z_{c}(4020)$ state would more likely make it a real hadron. The total absence of such peak would lean its interpretation to that of a dynamic accident.

\begin{figure}[h] \centering
\includegraphics[width=0.45\columnwidth]{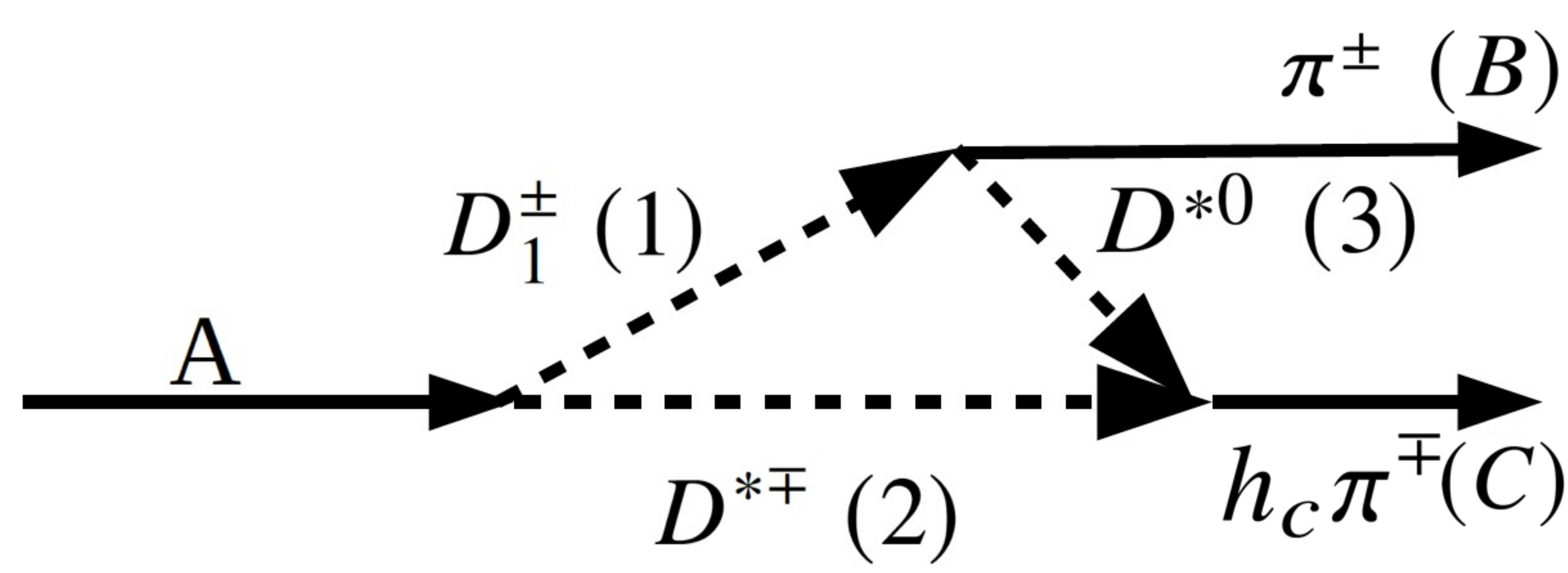}
\caption{Triangle diagram depicting the reaction shown in Eq.~(\ref{reaction}). 
\label{fig:data}}
\end{figure}

The calculation supporting that statement proceeds by considering the triangle singularity coming from the $(D^*_1D^*D^*)$ triangle loop (displayed in Fig.~\ref{fig:data}) satisfying the Coleman-Norton theorem, 
\begin{eqnarray}
 (A) \ \to \underbrace{D_1^{\pm}\ (1)}_{\to \pi^{\pm}\ (B)\ D^{*0}\ (3)} D^{*\mp}\ (2) \
  \xrightarrow{(2) + (3)} \ \pi^{\pm}\ (B)\ (h_c \pi ^{\mp })\ (C)  \ .
\label{reaction}
\end{eqnarray}
As discussed in~\cite{Liu:2014spa} and references therein, the invariant mass distribution of $ \pi h_c$ at an appropriate $E_{CM}$ gives rise to a peak playing the role of the $Z_{c}(4020)$. 
Analogously to the $Z_{c}(3900)$ structure investigated in~\cite{Abreu:2021xpz}, this is a typical reaction that should see the triangle singularity notably influenced by the medium in a heavy-ion collision, due to the fulfillment of the two alluded conditions.

First, the time of flight of the particles in the triangle (as computed in ~\cite{Abreu:2020jsl}) is $\tau_A \sim 9$ fm/c (with the values of $m_C = 4020$ MeV and $\Gamma_{D_1}= 31$ MeV), sufficiently short for the triangle reaction to often be completed in a relativistic heavy ion collision  (remarking that the duration of the
hadronic phase is of the order of 10 fm/c). And second, the masses and widths of the $D$-family mesons in the triangle diagram are sufficiently altered from their vacuum
values, as shown in Table~\ref{thermalmasses}.

\begin{center}
\begin{table}[h]
\caption{Meson thermal masses and widths (all in GeV)~\cite{Montana:2020vjg} used as input in the squared scalar triangle loop integral, defined in Eq. (\ref{loop-finiteT}), for the reaction in Eq.~(\ref{reaction}). 
We assumed $m_\pi$ to be constant (see Ref.~\cite{Abreu:2020jsl} for a detailed discussion). The mass of the initial state is fixed at $m_A = 4.430$ GeV. Since we are not aware of an evaluation of the $D_1$ thermal mass and width, we use an estimate according to \cite{Montana:2020vjg} for the excited states. \label{thermalmasses}
} 
\begin{center}
\begin{tabular}{|cccc|cccc|}
\hline
\hline
T & $ 0.00$  & $ 0.08$ & $ 0.15$   &  & $ 0.00$  & $ 0.08$ & $ 0.15$ \\ 
\hline
$m_{\pi} $        & 0.1396  & 0.1396 & 0.1396 & & & & \\
$m_{D^{\ast 0}} $ & 2.00685 &  2.0001 & 1.868 &$\Gamma_{D^{\ast 0}} $ & $55 \times 10^{-6}$ & 0.0022 & 0.071  \\
$m_{D^{\ast \pm }} $ & 2.0103 & 2.0024 & 1.872 & $\Gamma_{D^{\ast}}$ & $83.4 \times 10^{-6}$ & 0.0032 & 0.068  \\
$m_{D_1} $        & 2.4181  & 2.4134 & 2.4071 & $\Gamma_{D_1} $   & 0.0313    & 0.0325     & 0.040  \\
\hline \hline
\end{tabular}
\end{center}
\end{table}
\end{center}

The estimate of the change of the  triangle peak due to the medium influence can be obtained considering the scalar three-point loop integral in thermal field theory, 
\begin{eqnarray}    
\label{loop-finiteT} 
I_{\triangleleft}  \simeq    \frac{1}{2} \int \frac{d^3 q}{(2 \pi)^3} \frac{1}{8 E_1 E_2 E_3} \frac{1}{\left(E_A - \tilde E_1 -\tilde E_2 \right)} 
  \frac{1}{\left( E_C - \tilde E_2 - \tilde E_3 \right)}   \frac{1}{\left( E_B - \tilde E_1 + \tilde E_3 \right)}  \times \nonumber \\   
  \nonumber
  \left\{ \left[ 1 + 2n_\beta(\tilde E_2) \right] \left(\! \tilde E_1 \!   -\! E_B  \! -\! \tilde E_3\! \right) 
+
\left[ 1+2n_\beta(E_A- \tilde E_1 ) \right] \left(\! E_C \! -\!  \tilde E_2\! -\!  \tilde E_3\! \right)  \right.\\ \left.
+  \left[ 1+2n_\beta (\tilde E_3-E_C) \right] 
\left(\!E_A\! -\! \tilde E_1\! -\! \tilde E_2\! \right) \right\}\ .
\end{eqnarray}

\begin{figure}
\centering
\includegraphics[width=0.6\columnwidth]{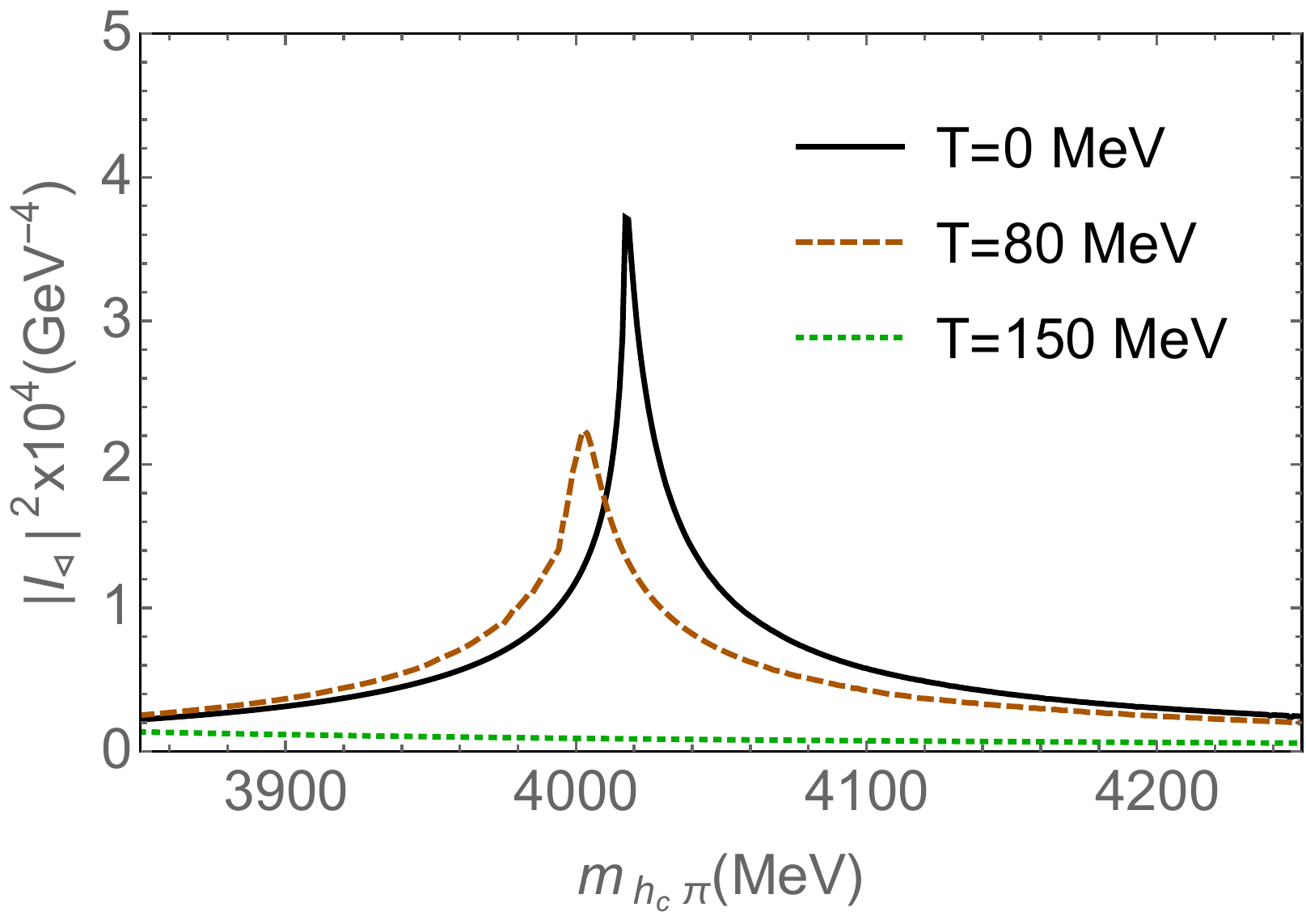}
\caption{\label{fig:Zc} 
Squared three-point loop integral $(|I_\triangleleft|^2)$ in Eq.~(\ref{loop-finiteT}) against $m_C$ for $(A) \rightarrow \pi^{\pm}\ (B)\ (h_c \pi ^{\mp })\ (C)$, at different temperatures. Meson thermal masses and widths used as input in the loop are shown in Table~\ref{thermalmasses}.}
\end{figure}

We plot in Fig.~\ref{fig:Zc} the squared three-point loop integral $(|I_\triangleleft|^2)$. It can be noticed a clear peak structure at zero temperature because of the triangle singularity. However, as the temperature increases the masses and widths of the $D$ mesons in the loop are modified, and the peak suffers a reduction of its height and a shift to smaller $ h_c \pi $ invariant mass, accounting for the decrease of the $D^{*} D^{*}$ threshold. At temperatures near the confinement/deconfinement phase transition, the peak structure is suppressed. Thus, as expected in light of our previous works, these findings indicate that this cusp is sensitive to the medium effects.

In the end, heavy-ion collisions might  shed light on the nature of this $Z_c(4020)$ and many other peaks, noticing that it is an ambience where the triangle singularity does not play an important role.

\section*{Acknowledgement}
\noindent
This work was supported by spanish  MICINN PID2019-108655GB-I00 and -106080GB-C21; EU’s 824093 (STRONG2020); and UCM’s 910309 group grants and IPARCOS, as well as the Brazilian CNPq (contracts 309950/2020-1 and 400546/2016-7) and FAPESB (contract
INT0007/2016).


\end{document}